\pdfoutput=1

\documentclass[12pt,a4paper]{article}

\usepackage{ifthen} 
\newboolean{pdflatex}
\setboolean{pdflatex}{true} 

\newboolean{articletitles}
\setboolean{articletitles}{true} 

\newboolean{uprightparticles}
\setboolean{uprightparticles}{false} 

\newboolean{inbibliography}
\setboolean{inbibliography}{false} 

\def\paperauthors{LHCb collaboration} 
\def\paperasciititle{Observation of a new Xib- resonance} 
\def\papertitle{Observation of a new $\Xibm$ resonance} 
\def\paperkeywords{{High Energy Physics}, {LHCb}} 
\def\papercopyright{\the\year\ CERN for the benefit of the LHCb collaboration}
\def\paperlicence{CC-BY-4.0}
\def\paperlicenceurl{https://creativecommons.org/licenses/by/4.0/}

\usepackage[top=1in, bottom=1.25in, left=1in, right=1in]{geometry}

\usepackage{microtype}
\usepackage{lineno}  
\usepackage{xspace} 
\usepackage{caption} 

\usepackage{graphicx}  
\usepackage{color}
\usepackage{colortbl}
\graphicspath{{./figs/}} 

\usepackage{amsmath} 
\usepackage{amssymb}
\usepackage{amsfonts}
\usepackage{upgreek} 

\newcommand*\patchAmsMathEnvironmentForLineno[1]{%
\expandafter\let\csname old#1\expandafter\endcsname\csname #1\endcsname
\expandafter\let\csname oldend#1\expandafter\endcsname\csname
end#1\endcsname
 \renewenvironment{#1}%
   {\linenomath\csname old#1\endcsname}%
   {\csname oldend#1\endcsname\endlinenomath}%
}
\newcommand*\patchBothAmsMathEnvironmentsForLineno[1]{%
  \patchAmsMathEnvironmentForLineno{#1}%
  \patchAmsMathEnvironmentForLineno{#1*}%
}
\AtBeginDocument{%
\patchBothAmsMathEnvironmentsForLineno{equation}%
\patchBothAmsMathEnvironmentsForLineno{align}%
\patchBothAmsMathEnvironmentsForLineno{flalign}%
\patchBothAmsMathEnvironmentsForLineno{alignat}%
\patchBothAmsMathEnvironmentsForLineno{gather}%
\patchBothAmsMathEnvironmentsForLineno{multline}%
\patchBothAmsMathEnvironmentsForLineno{eqnarray}%
}

\usepackage{hyperxmp}

\usepackage[pdftex,
            pdfauthor={\paperauthors},
            pdftitle={\paperasciititle},
            pdfkeywords={\paperkeywords},
            pdfcopyright={Copyright (C) \papercopyright},
            pdflicenseurl={\paperlicenceurl}]{hyperref}

\usepackage[all]{hypcap} 

\usepackage{xspace} 
\usepackage{upgreek}

\def\lhcb {\mbox{LHCb}\xspace}

\def\MagUp {\mbox{\em Mag\kern -0.05em Up}\xspace}

\ifthenelse{\boolean{uprightparticles}}%
{

 \def\Pmu         {\ensuremath{\upmu}\xspace}

 \def\Ppi         {\ensuremath{\uppi}\xspace}

 \def\Ptau        {\ensuremath{\uptau}\xspace}

 \def\Ppsi        {\ensuremath{\uppsi}\xspace}

 \def\PDelta      {\ensuremath{\Delta}\xspace}                 
 \def\PXi      {\ensuremath{\Xi}\xspace}                 
 \def\PLambda      {\ensuremath{\Lambda}\xspace}                 
 \def\PSigma      {\ensuremath{\Sigma}\xspace}                 
 \def\POmega      {\ensuremath{\Omega}\xspace}                 
 \def\PUpsilon      {\ensuremath{\Upsilon}\xspace}                 
 
 \def\PB      {\ensuremath{\mathrm{B}}\xspace}                 
                  
 \def\PD      {\ensuremath{\mathrm{D}}\xspace}

 \def\PJ      {\ensuremath{\mathrm{J}}\xspace}                 
 \def\PK      {\ensuremath{\mathrm{K}}\xspace}

 \def\Pb      {\ensuremath{\mathrm{b}}\xspace}                 
 \def\Pc      {\ensuremath{\mathrm{c}}\xspace}

 \def\Pi      {\ensuremath{\mathrm{i}}\xspace}

 \def\Ps      {\ensuremath{\mathrm{s}}\xspace}

}
{

 \def\Pmu         {\ensuremath{\mu}\xspace}

 \def\Ppi         {\ensuremath{\pi}\xspace}

 \def\Ptau        {\ensuremath{\tau}\xspace}

 \def\Ppsi        {\ensuremath{\psi}\xspace}                 
                  
 \mathchardef\PDelta="7101
 \mathchardef\PXi="7104
 \mathchardef\PLambda="7103
 \mathchardef\PSigma="7106
 \mathchardef\POmega="710A
 \mathchardef\PUpsilon="7107
                  
 \def\PB      {\ensuremath{B}\xspace}                 
                  
 \def\PD      {\ensuremath{D}\xspace}

 \def\PJ      {\ensuremath{J}\xspace}                 
 \def\PK      {\ensuremath{K}\xspace}

 \def\Pb      {\ensuremath{b}\xspace}                 
 \def\Pc      {\ensuremath{c}\xspace}

 \def\Pi      {\ensuremath{i}\xspace}

 \def\Ps      {\ensuremath{s}\xspace}

}

\makeatletter
\ifcase \@ptsize \relax
  \newcommand{\miniscule}{\@setfontsize\miniscule{4}{5}}
\or
  \newcommand{\miniscule}{\@setfontsize\miniscule{5}{6}}
\or
  \newcommand{\miniscule}{\@setfontsize\miniscule{5}{6}}
\fi
\makeatother

\DeclareRobustCommand{\optbar}[1]{\shortstack{{\miniscule (\rule[.5ex]{1.25em}{.18mm})}
  \\ [-.7ex] $#1$}}

\def\mup        {{\ensuremath{\Pmu^+}}\xspace}
\def\mun        {{\ensuremath{\Pmu^-}}\xspace} 

\def\taum       {{\ensuremath{\Ptau^-}}\xspace}

\def\squark    {{\ensuremath{\Ps}}\xspace}

\def\cquark    {{\ensuremath{\Pc}}\xspace}

\def\bquark    {{\ensuremath{\Pb}}\xspace}

\def\pion   {{\ensuremath{\Ppi}}\xspace}

\def\pip    {{\ensuremath{\pion^+}}\xspace}
\def\pim    {{\ensuremath{\pion^-}}\xspace}

\def\kaon    {{\ensuremath{\PK}}\xspace}
  \def\Kbar    {{\kern 0.2em\overline{\kern -0.2em \PK}{}}\xspace}

\def\KorKbar    {\kern 0.18em\optbar{\kern -0.18em K}{}\xspace}

\def\Kp      {{\ensuremath{\kaon^+}}\xspace}
\def\Km      {{\ensuremath{\kaon^-}}\xspace}

  \def\Dbar    {{\kern 0.2em\overline{\kern -0.2em \PD}{}}\xspace}
\def\D       {{\ensuremath{\PD}}\xspace}

\def\DorDbar    {\kern 0.18em\optbar{\kern -0.18em D}{}\xspace}
\def\Dz      {{\ensuremath{\D^0}}\xspace}

\def\Dp      {{\ensuremath{\D^+}}\xspace}

\def\Dstarp  {{\ensuremath{\D^{*+}}}\xspace}

\def\Dsp     {{\ensuremath{\D^+_\squark}}\xspace}

\def\Bbar    {{\ensuremath{\kern 0.18em\overline{\kern -0.18em \PB}{}}}\xspace}

\def\BorBbar    {\kern 0.18em\optbar{\kern -0.18em B}{}\xspace}

\def\jpsi     {{\ensuremath{{\PJ\mskip -3mu/\mskip -2mu\Ppsi\mskip 2mu}}}\xspace}

  \def\Y#1S{\ensuremath{\PUpsilon{(#1S)}}\xspace}

\def\Xires       {{\ensuremath{\PXi}}\xspace}

\def\Lz          {{\ensuremath{\PLambda}}\xspace}
\def\Lbar        {{\ensuremath{\kern 0.1em\overline{\kern -0.1em\PLambda}}}\xspace}
\def\LorLbar    {\kern 0.18em\optbar{\kern -0.18em \PLambda}{}\xspace}

\def\XibStar     {{\ensuremath{\Xires_\bquark(6227)^{-}}}\xspace}
\def\Lb      {{\ensuremath{\Lz^0_\bquark}}\xspace}

\def\Lc      {{\ensuremath{\Lz^+_\cquark}}\xspace}

\def\Xib     {{\ensuremath{\Xires_\bquark}}\xspace}
\def\Xibz    {{\ensuremath{\Xires^0_\bquark}}\xspace}
\def\Xibm    {{\ensuremath{\Xires^-_\bquark}}\xspace}

\def\Xic     {{\ensuremath{\Xires_\cquark}}\xspace}
\def\Xicz    {{\ensuremath{\Xires^0_\cquark}}\xspace}
\def\Xicp    {{\ensuremath{\Xires^+_\cquark}}\xspace}

\def\BF         {{\ensuremath{\mathcal{B}}}\xspace}

\def\BR         {\BF}

\def\to                 {\ensuremath{\rightarrow}\xspace}

\def\AT#1     {\ensuremath{A_{\mathrm{T}}^{#1}}\xspace}           

\def\C#1      {\ensuremath{\mathcal{C}_{#1}}\xspace}                       
\def\Cp#1     {\ensuremath{\mathcal{C}_{#1}^{'}}\xspace}                    
\def\Ceff#1   {\ensuremath{\mathcal{C}_{#1}^{\mathrm{(eff)}}}\xspace}        
\def\Cpeff#1  {\ensuremath{\mathcal{C}_{#1}^{'\mathrm{(eff)}}}\xspace}       
\def\Ope#1    {\ensuremath{\mathcal{O}_{#1}}\xspace}                       
\def\Opep#1   {\ensuremath{\mathcal{O}_{#1}^{'}}\xspace}                    

\newcommand{\tev}{\ifthenelse{\boolean{inbibliography}}{\ensuremath{~T\kern -0.05em eV}}{\ensuremath{\mathrm{\,Te\kern -0.1em V}}}\xspace}
\newcommand{\gev}{\ensuremath{\mathrm{\,Ge\kern -0.1em V}}\xspace}
\newcommand{\mev}{\ensuremath{\mathrm{\,Me\kern -0.1em V}}\xspace}
\newcommand{\kev}{\ensuremath{\mathrm{\,ke\kern -0.1em V}}\xspace}
\newcommand{\ev}{\ensuremath{\mathrm{\,e\kern -0.1em V}}\xspace}
\newcommand{\gevc}{\ensuremath{{\mathrm{\,Ge\kern -0.1em V\!/}c}}\xspace}
\newcommand{\mevc}{\ensuremath{{\mathrm{\,Me\kern -0.1em V\!/}c}}\xspace}
\newcommand{\gevcc}{\ensuremath{{\mathrm{\,Ge\kern -0.1em V\!/}c^2}}\xspace}
\newcommand{\gevgevcccc}{\ensuremath{{\mathrm{\,Ge\kern -0.1em V^2\!/}c^4}}\xspace}
\newcommand{\mevcc}{\ensuremath{{\mathrm{\,Me\kern -0.1em V\!/}c^2}}\xspace}

\def\invfb   {\ensuremath{\mbox{\,fb}^{-1}}\xspace}

\newcommand{\stat}{\ensuremath{\mathrm{\,(stat)}}\xspace}
\newcommand{\syst}{\ensuremath{\mathrm{\,(syst)}}\xspace}

\newcommand{\chisq}{\ensuremath{\chi^2}\xspace}

\newcommand{\chisqip}{\ensuremath{\chi^2_{\text{IP}}}\xspace}

\def\gsim{{~\raise.15em\hbox{$>$}\kern-.85em
          \lower.35em\hbox{$\sim$}~}\xspace}
\def\lsim{{~\raise.15em\hbox{$<$}\kern-.85em
          \lower.35em\hbox{$\sim$}~}\xspace}

\def\ptot       {\mbox{$p$}\xspace}
\def\pt         {\mbox{$p_{\mathrm{ T}}$}\xspace}

\def\tell1  {TELL1\xspace}
\def\ukl1   {UKL1\xspace}

\usepackage{cite} 
\usepackage{mciteplus}

\usepackage{longtable} 

\begin{document}

\renewcommand{\thefootnote}{\fnsymbol{footnote}}
\setcounter{footnote}{1}


\begin{titlepage}
\pagenumbering{roman}

\vspace*{-1.5cm}
\centerline{\large EUROPEAN ORGANIZATION FOR NUCLEAR RESEARCH (CERN)}
\vspace*{1.5cm}
\noindent
\begin{tabular*}{\linewidth}{lc@{\extracolsep{\fill}}r@{\extracolsep{0pt}}}
\ifthenelse{\boolean{pdflatex}}
{\vspace*{-1.5cm}\mbox{\!\!\!\includegraphics[width=.14\textwidth]{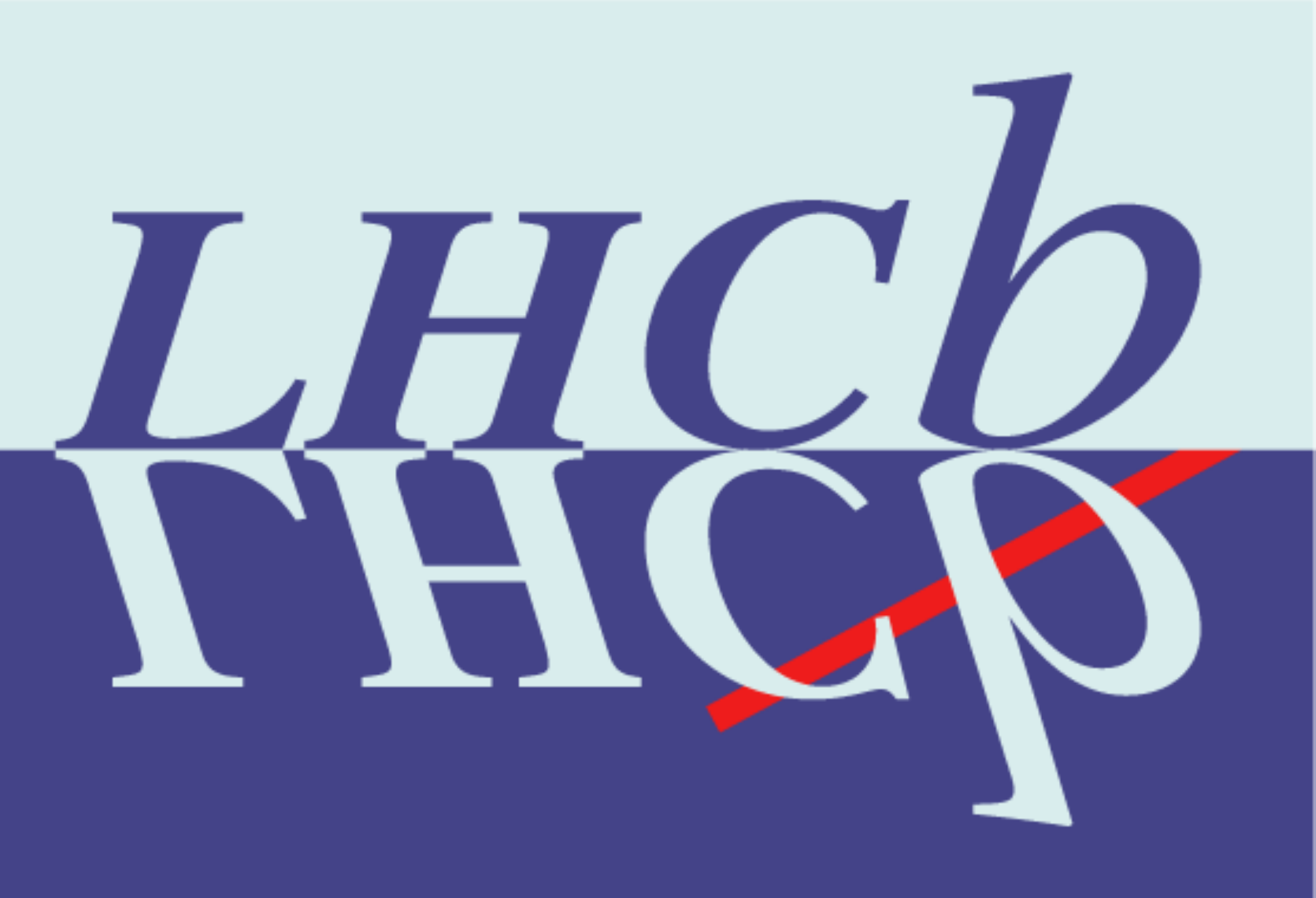}} & &}%
{\vspace*{-1.2cm}\mbox{\!\!\!\includegraphics[width=.12\textwidth]{lhcb-logo.eps}} & &}%
\\
 & & CERN-EP-2018-108 \\  
 & & LHCb-PAPER-2018-013 \\  
 & & May 23, 2018 \\ 
 & & \\
\end{tabular*}

\vspace*{4.0cm}

{\normalfont\bfseries\boldmath\huge
\begin{center}
  \papertitle 
\end{center}
}

\vspace*{2.0cm}

\begin{center}
\paperauthors\footnote{Authors are listed at the end of this paper.}
\end{center}

\vspace{\fill}

\begin{abstract}
  \noindent
  From samples of $pp$ collision data collected by the LHCb experiment
at $\sqrt{s}=7$, $8$ and $13$\tev, corresponding to integrated luminosities of 
1.0, 2.0 and 1.5\invfb, respectively, a peak
in both the $\Lb\Km$ and $\Xibz\pim$ invariant mass spectra is observed. In the quark model, radially and orbitally excited 
$\Xibm$ resonances with quark content $bds$ are expected. Referring to this peak as
$\XibStar$, the mass and natural width are measured to be 
$m_{\XibStar}=6226.9\pm2.0\pm0.3\pm0.2$\mevcc and $\Gamma_{\XibStar}=18.1\pm5.4\pm1.8$\mevcc, where the first uncertainty
is statistical, the second is systematic, and the third, on $m_{\XibStar}$, is due to the knowledge of the $\Lb$ baryon 
mass.
Relative production rates of the ${\XibStar\to\Lb\Km}$ and ${\XibStar\to\Xibz\pim}$ decays are also reported.
\end{abstract}

\vspace*{2.0cm}

\begin{center}
  Published in Phys.~Rev.~Lett. 121 (2018) 072002
\end{center}

\vspace{\fill}

{\footnotesize 
\centerline{\copyright~\papercopyright, licence \href{\paperlicenceurl}{\paperlicence}.}}
\vspace*{2mm}

\end{titlepage}


\newpage
\setcounter{page}{2}
\mbox{~}

\cleardoublepage


\renewcommand{\thefootnote}{\arabic{footnote}}
\setcounter{footnote}{0}



\pagestyle{plain} 
\setcounter{page}{1}
\pagenumbering{arabic}


%

In the constituent quark model~\cite{GellMann:1964nj,Zweig:1964jf}, baryonic states
form multiplets according to the symmetry of their flavor, spin, and spatial wave functions.
The masses, widths and decay modes of these states give insight into their internal structure~\cite{Klempt:2009pi}. 
The $\Xibz$ and $\Xibm$ states form an isodoublet of $bsq$ bound states, where
$q$ is a $u$ or $d$ quark, respectively. Three such isodoublets, which are neither radially nor orbitally excited, should exist~\cite{Ebert:1995fp}, 
and include one with spin $j_{qs}=0$ and $J^P=(1/2)^+$ ($\Xib$), a second with $j_{qs}=1$ and $J^P=(1/2)^+$ ($\Xib^{\prime}$),
and a third with $j_{qs}=1$ and $J^P=(3/2)^+$ ($\Xib^{*}$). Here, $j_{qs}$ is the spin of the light diquark system 
$qs$, and $J^P$ represents the spin and parity of the state. Three of the four $j_{qs}=1$ states
have been recently observed through
their decays to $\Xibz\pim$ and $\Xibm\pip$~\cite{LHCb-PAPER-2014-061,Chatrchyan:2012ni,LHCb-PAPER-2016-010}. 

Beyond these lowest-lying states, a spectrum of heavier states is 
expected~\cite{Ebert:2011kk,Ebert:2007nw,Roberts:2007ni,Garcilazo:2007eh,Chen:2014nyo,Mao:2015gya,Grach:2008ij,PhysRevD.87.034032,Karliner:2008sv,Wang:2010it,Valcarce:2008dr,Vijande:2012mk,Wang:2017kfr,Wang:2017goq,Chen:2016phw,Thakkar:2016dna},
where there are either radial or orbital excitations amongst the constituent quarks.
The only such states discovered thus far in the $b$-baryon sector are the $\Lz_b(5912)^0$ 
and $\Lz_b(5920)^0$ resonances~\cite{LHCb-PAPER-2012-012}, which are consistent with
being orbital excitations of the $\Lb$ baryon. 

In this Letter, we report the first observation of a new state, decaying into both
$\Lb\Km$ and $\Xibz\pim$, using samples of $pp$ collision data collected with the \lhcb experiment
at 7, 8 and 13\tev,
corresponding to integrated luminosities of 1.0, 2.0 and 1.5\invfb, respectively.
The observation of these decays is consistent with the strong decay of a radially or 
orbitally excited $\Xibm$ baryon, hereafter referred 
to as $\XibStar$. Charge-conjugate processes are implicitly included throughout this Letter.

The mass and width of the $\XibStar$ baryon are measured using the $\Lb\Km$ mode, where the
$\Lb$ baryon is detected through its fully reconstructed hadronic (HAD) decay to $\Lc\pim$. 
Larger samples of semileptonic (SL) $\Lb$ and $\Xibz$ decays are used to measure the production ratios 
\begin{align}
R(\Lb\Km) &\equiv \frac{f_{\XibStar}}{f_{\Lb}}\BF(\XibStar\to\Lb\Km), \\
R(\Xibz\pim) &\equiv \frac{f_{\XibStar}}{f_{\Xibz}}\BF(\XibStar\to\Xibz\pim),
\end{align}
\noindent where $f_{\XibStar}$, $f_{\Xibz}$ and $f_{\Lb}$ are the fragmentation fractions of a $b$ quark into each
baryon and $\BF$ represents a branching fraction. Here, the $\Lb$ and $\Xibz$ baryons are detected using
$\Lb\to\Lc\mun X$ and $\Xibz\to\Xicp\mun X$ decays, where $X$ represents undetected particles.
Throughout the text, $H_b^0$ ($H_c^+$) is used to designate either a $\Lb$ or $\Xibz$ ($\Lc$ or $\Xicp$) baryon.
Owing to much larger branching fractions, the SL signal yields are about an order of magnitude larger than that of any
fully hadronic final state, which enables the observation of the $\XibStar\to\Xibz\pim$ mode. The SL decays are not 
used in the $\XibStar$ mass or width determination, as they have larger systematic uncertainties due to modeling of
the mass resolution.

The \lhcb detector~\cite{Alves:2008zz,LHCb-DP-2014-002} is a single-arm forward
spectrometer covering the \mbox{pseudorapidity} range $2<\eta <5$,
designed for the study of particles containing \bquark or \cquark
quarks~\cite{Alves:2008zz,LHCb-DP-2014-002}. 

The tracking system provides a measurement of the momentum, \ptot, of charged particles with
a relative uncertainty that varies from 0.5\% at low momentum to 1.0\% at 200\gevc.
Events are selected online by a trigger,
which consists of a hardware stage, based on information from the calorimeter and muon
systems, followed by a software stage, which applies a full event reconstruction~\cite{LHCb-DP-2012-004,BBDT}.
Simulated data samples are produced using the software packages described in 
Refs.~\cite{Sjostrand:2007gs,Sjostrand:2006za,LHCb-PROC-2010-056,Lange:2001uf,Golonka:2005pn,Allison:2006ve, *Agostinelli:2002hh,LHCb-PROC-2011-006}.

Samples of $\Lb$ ($\Xibz$) are formed from $\Lc\pim$ and $\Lc\mun$ ($\Xicp\mun$)
combinations, where $\Lc$ and $\Xicp$ decays are reconstructed in the $p\Km\pip$ final
state. The $H_c^+$ decay products must have particle identification (PID) information consistent with the 
given particle hypothesis, and be inconsistent with originating from a primary vertex (PV)
by requiring each to have large $\chisqip$ with respect to all PVs in the event.
Here $\chisqip$ is the difference in $\chisq$ of the vertex fit of a given PV when the particle (here $p$, $\Km$ or $\pip$) 
is included or excluded from the fit. The $H_c^+$ candidate must have a fitted vertex significantly
displaced from all PVs in the event and have an invariant mass within 60\mevcc of the known $H_c^+$ mass.

The $H_c^+$ background is dominated by random combinations of tracks from nonsignal $b$-hadron decays.
In the $\Xicp$ sample, about 15\% of this background is due to misidentified
$\Dp\to\Km\pip\pip,~\Dp\to\Kp\Km\pip$, $\Dsp\to\Kp\Km\pip$ and $\Dstarp\to(\Dz\to\Km\pip)\pip$ decays.
These cross-feed contributions are suppressed by employing tighter PID requirements on candidates that are consistent
with being one of these charm mesons, with only a 1\% loss of signal efficiency. These tighter requirements are not applied
to the $\Lc$ sample, as the cross-feed contributions are negligible.

Muon (pion) candidates with transverse momentum $\pt>1\gevc$ ($0.5$\gevc) and large $\chisqip$ 
are combined with $H_c^+$ candidates to form the $H_b^0$ samples.
Each $H_b^0$ decay vertex is required to be significantly displaced from all PVs in the event. 
For the $\Lb\to\Lc\pim$ decay, the reconstructed $\Lb$ trajectory must point back to one of the
PVs in the event; only a very loose pointing requirement is imposed on the SL decay due to the momentum
carried by the undetected particles. To reduce background in the SL decay samples, 
the $z$ coordinates of the $H_c^+$ and $H_b^0$ decay vertices are required to satisfy $z(H_c^+)-z(H_b^0)>-0.05$~mm, where
$z$ is measured along the beam direction. Candidates that satisfy the invariant mass requirements, 
$5.2<M(\Lc\pim)<6.0$\gevcc or $M(H_c^+\mun)<8$\gevcc, are retained, where
$M$ designates the invariant mass of the system of indicated particle(s).

To further suppress background in the $\Xibz\to\Xicp\mun X$ sample, a boosted decision tree (BDT) 
discriminant~\cite{Roe,AdaBoost} is used. The BDT exploits fourteen input variables:
the $\chisq$ values of the fitted $\Xicp$ and $\Xibz$ 
decay vertices, and the momentum, $\pt$, $\chisqip$ and a PID variable for each $\Xicp$ final-state particle.
Simulated signal decays and background from the $\Xicp$ mass sidebands,
$30<|M(p\Km\pip)-m_{\Xicp}|<60$\mevcc, in data are used to train the BDT, where
$m$ refers to the known mass of the indicated particle~\cite{PDG2017}. The PID response for final-state hadrons in
the signal decay is obtained from large $\Lz\to p\pim$ and $\Dstarp\to(\Dz\to\Km\pip)\pip$ calibration samples in data, which 
is weighted to reproduce the kinematics of the signal.
The chosen requirement on the BDT response provides an efficiency of about 90\% (40\%) on the signal (background). 

\begin{figure}[tb]
\centering
\includegraphics[width=0.48\textwidth]{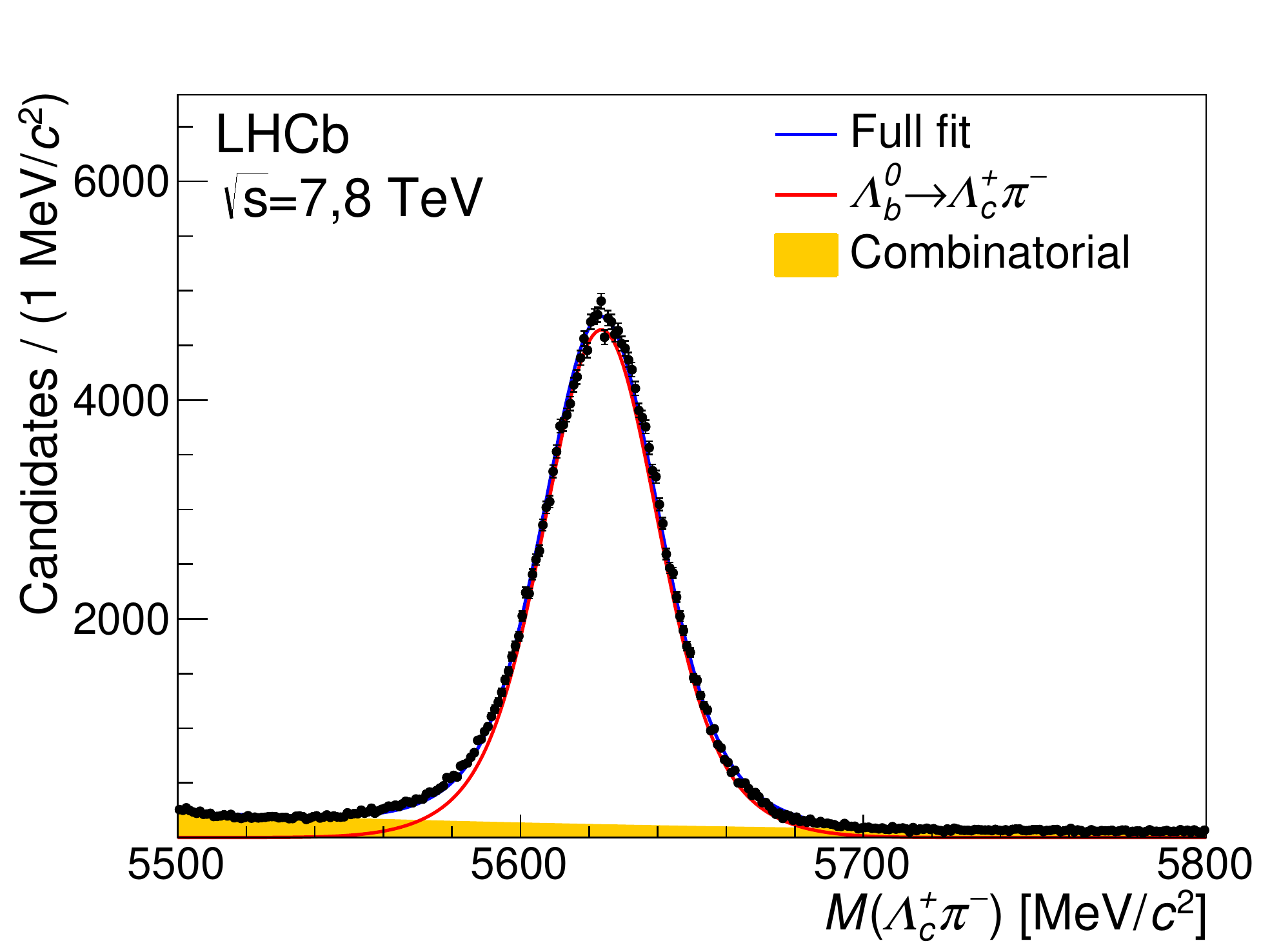}
\includegraphics[width=0.48\textwidth]{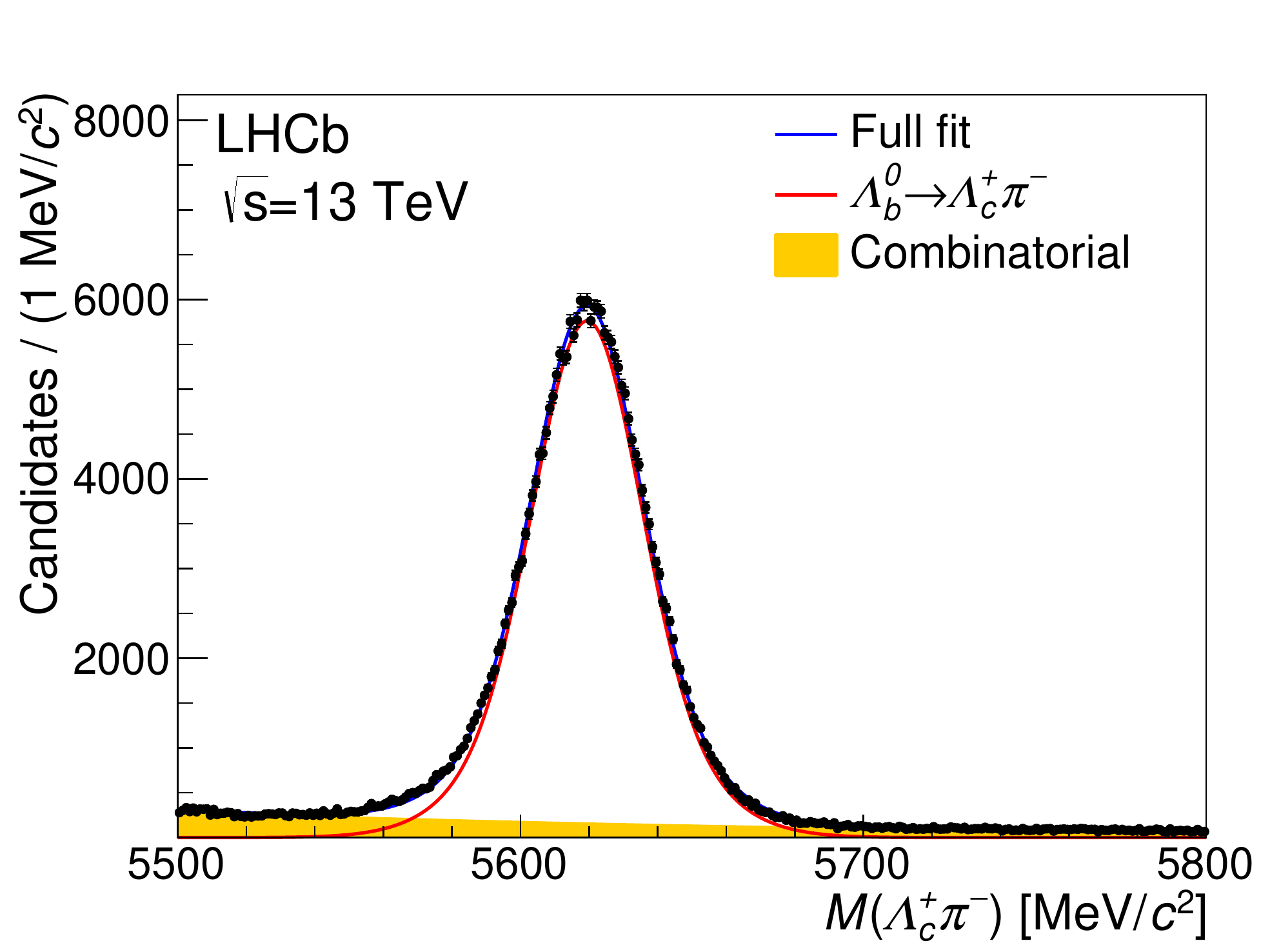}
\includegraphics[width=0.48\textwidth]{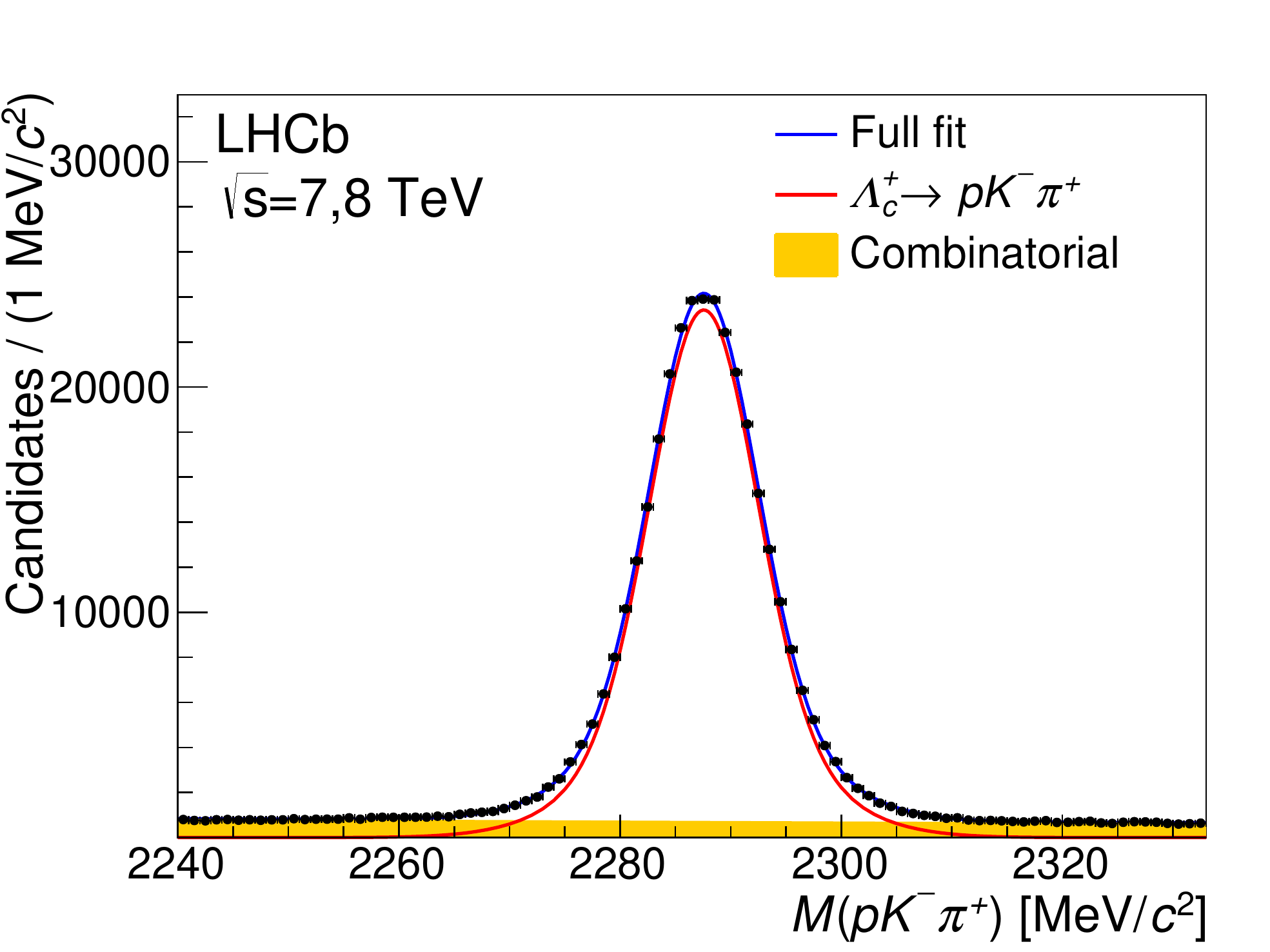}
\includegraphics[width=0.48\textwidth]{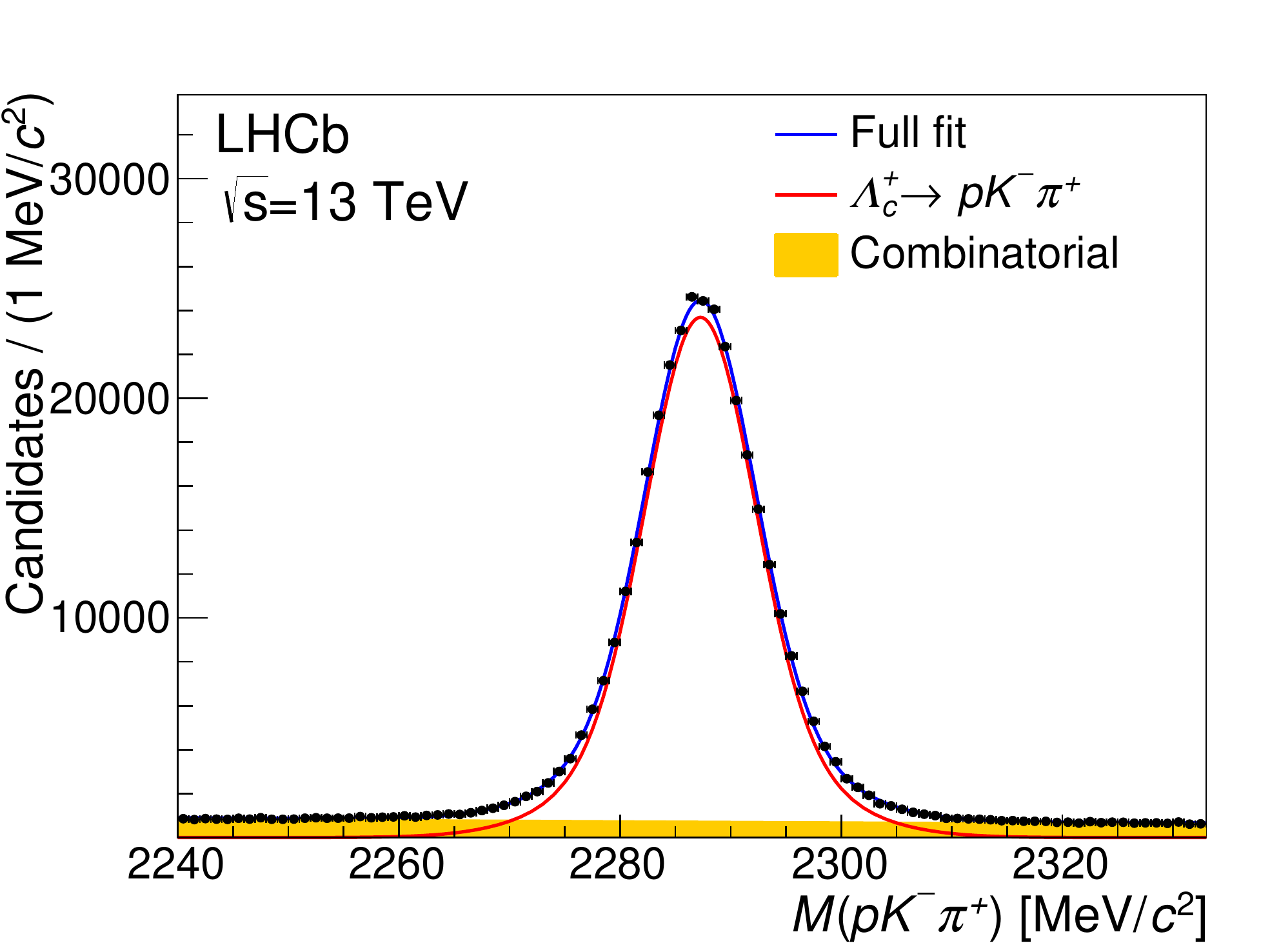}
\includegraphics[width=0.48\textwidth]{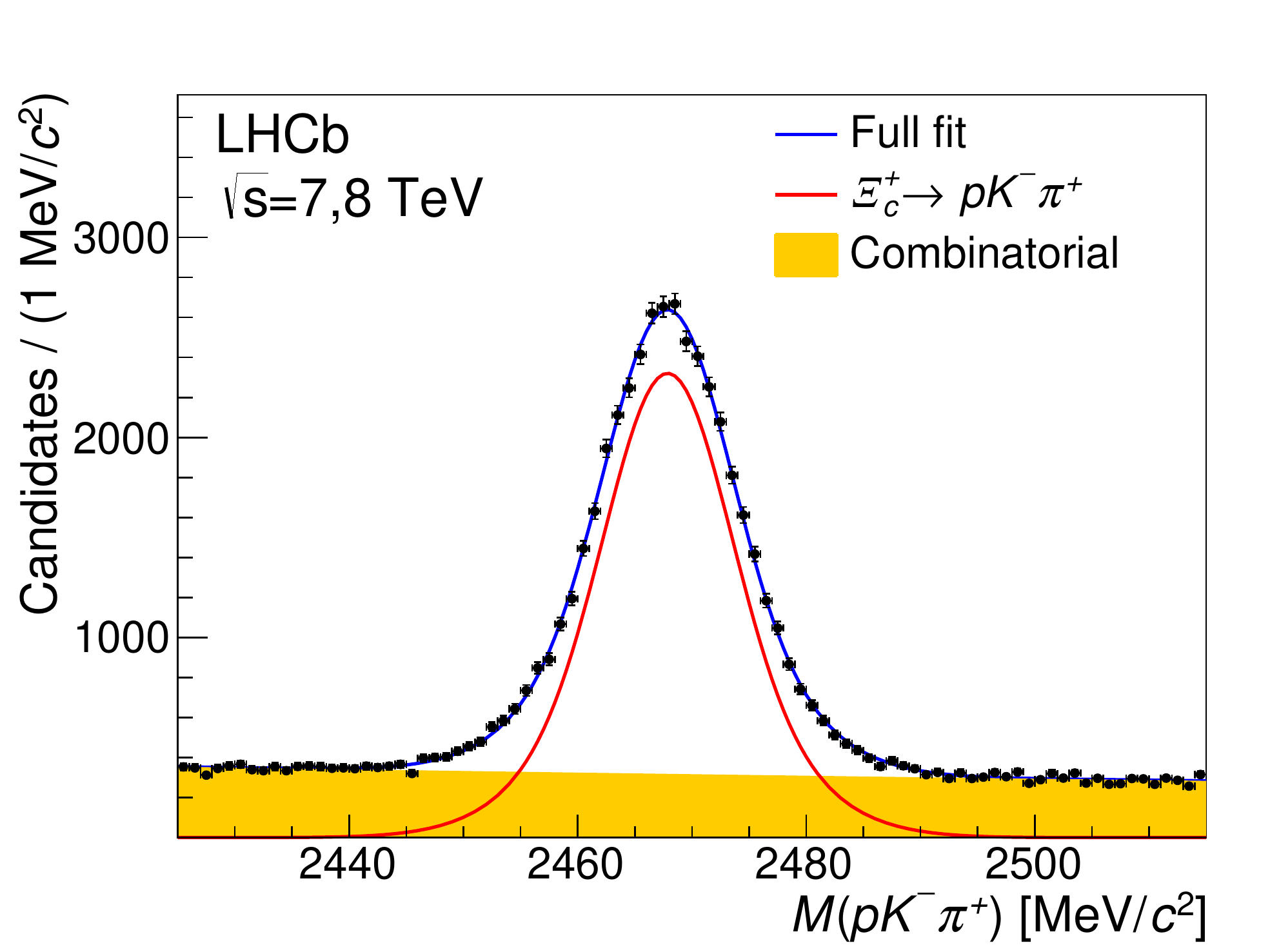}
\includegraphics[width=0.48\textwidth]{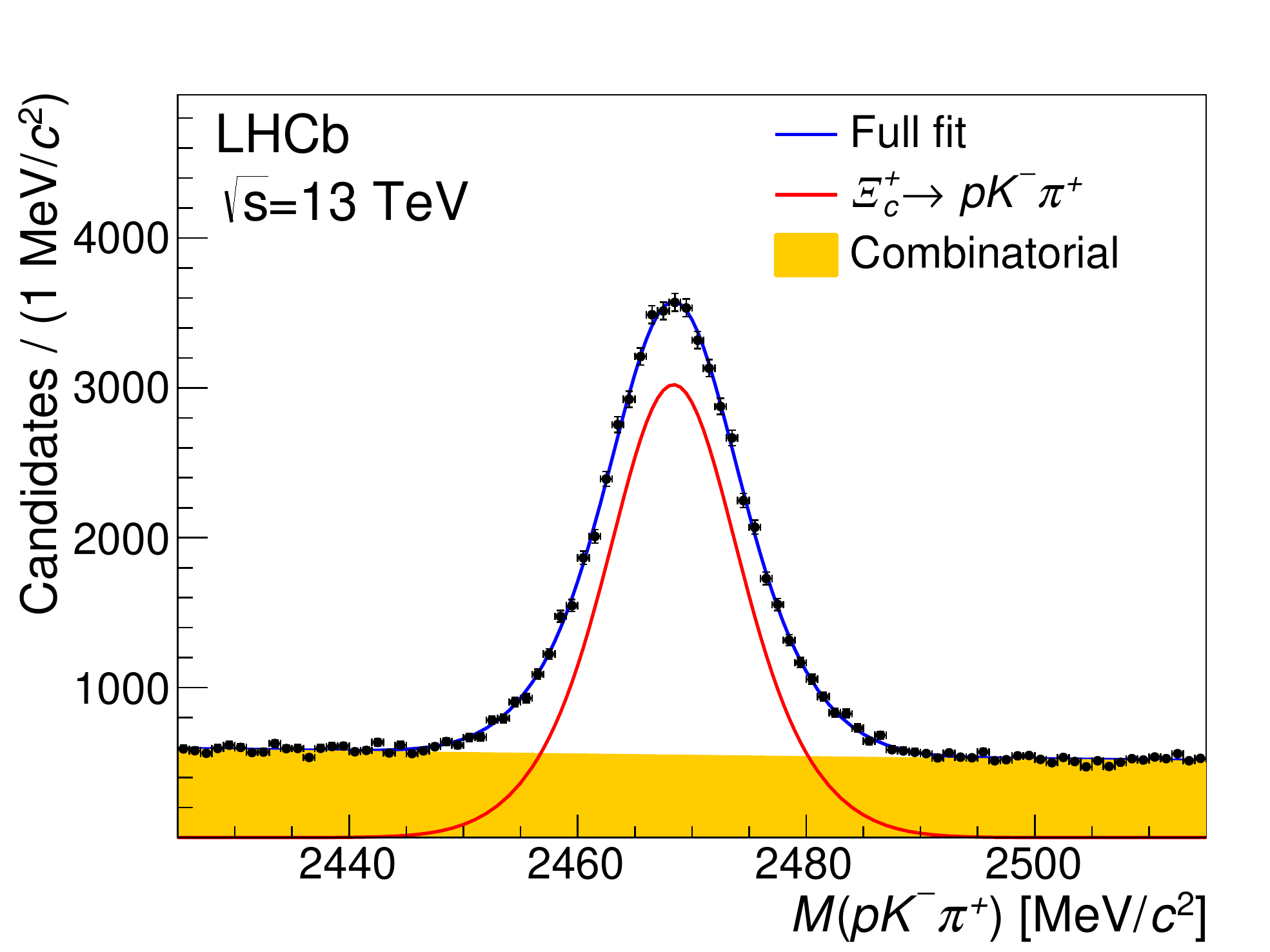}
\caption{\small{Invariant mass spectra for (top) $\Lb\to\Lc\pim$, (middle) $\Lc$ from 
$\Lb\to\Lc\mun X$, and (bottom)  $\Xicp$ from $\Xibz\to\Xicp\mun X$ candidate decays.
The left column is for $7,\,8$~TeV and the right is for 13~TeV data. 
Fits are overlaid, as described in the text.
Here, the $\Lb\to\Lc\mun X$ mode has been prescaled by a factor of ten.}}
\label{fig:NormModeMassDistributions}
\end{figure}

Figure~\ref{fig:NormModeMassDistributions} shows the mass spectra for 
$\Lb\to\Lc\pim$, $\Lc\to p\Km\pip$ (from $\Lb\to\Lc\mun X$) and $\Xicp\to p\Km\pip$ 
(from $\Xibz\to\Xicp\mun X$) candidates. For the $\Lb\to\Lc\pim$ mode, a peak at the known
$\Lb$ mass is seen. For the SL modes, the $\Lc$ and $\Xicp$ mass peaks are used to determine the number 
of $\Lb$ and $\Xibz$ baryon decays, as the combinatorial
background from random $H_c^+\mun$ combinations is at the 1\% level.
The mass spectra are fit with the sum of two Gaussian functions with a common mean to represent the signal component 
and an exponential background function. The yields are given in Table~\ref{tab:sumYield}.

\begin{table*}[tb]
\begin{center}
\caption{\small{Uncorrected $\XibStar$ and $H_b^0$ signal yields for $7,\,8$ and 13~TeV data. The $H_b^0$ yields 
are limited to the signal regions used to form $\XibStar$ candidates (see text).}} 
\begin{tabular}{lcccc}
\hline\hline
$\XibStar$              &                  \multicolumn{2}{c}{$7,\,8$\tev}                   &       \multicolumn{2}{c}{13\tev} \\
final state             & $N(\XibStar)$& $~~N(H_b^0)$ [$10^3$] &  $N(\XibStar)$   &     $~~N(H_b^0)$ [$10^3$] \\
\hline
$(\Lb)_{\rm HAD}\Km$   & $170\pm53$   &  $~\,204.6\pm0.5$    &  $215\pm63$      &   $~\,252.7\pm0.6$ \\
$(\Lb)_{\rm SL}\Km$    & $2772\pm325$ &  $3133\pm6$       &  $3701\pm432$    &   $3226\pm6$    \\
$(\Xibz)_{\rm SL}\pim$ & $351\pm68$   &  $~~~36.6\pm0.3$     &  $274\pm73$      &   $~~~46.5\pm0.3$ \\
\hline\hline
\end{tabular}
\label{tab:sumYield}
\end{center}
\end{table*}

To form $\XibStar$ candidates, a $\Lb$ ($\Xibz$) candidate is combined with a $\Km$ ($\pim$) meson that
has small $\chisqip$, consistent with being produced in the strong decay of the $\XibStar$ resonance.
Only $H_b^0$ candidates satisfying $|M(\Lc\pim)_{\rm HAD}-m_{\Lb}|<60$\mevcc, $|M(p\Km\pip)_{\rm SL}-m_{\Lc}|<15$\mevcc, and 
$|M(p\Km\pip)_{\rm SL}-m_{\Xicp}|<18$\mevcc are considered, where HAD and SL indicate the sample from which the mass
is determined. We require $p_{\rm T}^{\Km}>800$\mevc and $p_{\rm T}^{\pim}>900$\mevc, based on an optimization of the expected
statistical uncertainty on the $\XibStar$ signal yield, using simulation to model the signal and 
either wrong-sign ($\Lb\Kp,~\Xibz\pip$) or $\XibStar$ mass sideband samples in data to model the background.
After all selections the dominant source of background is due to combinations of real $\Lb$ ($\Xibz$) decays
with a random $\Km$ ($\pim$) meson. All candidates satisfying these selections are retained.

To improve the resolution on the $\XibStar$ mass, we use the mass differences 
${\delta m_K\equiv M(\Lb\Km)-M(\Lb)}$ and ${\delta m_{\pi}\equiv M(\Xibz\pim)-M(\Xibz)}$, for the $\Lb\Km$ and $\Xibz\pim$
final states, respectively. 
The $\delta m_{K(\pi)}$ resolution is obtained from simulated $\XibStar$ decays, where the decay width is set to a negligible value.
For the $\Lb\to\Lc\pim$ mode, the $\delta m_K$ resolution model is approximately Gaussian with a width of 2.4\mevcc.
For the SL decays, the missing momentum, $p_{\rm miss}$, is estimated by assuming it 
is carried by a zero-mass particle that balances the momentum transverse to the $H_b^0$ direction
(formed from its decay vertex and PV), and satisfies the mass constraint $(p_{H_c^+}+p_{\mun}+p_{\rm miss})^2=m_{H_b^0}^2$.
Mass resolution shape parameters are obtained by fitting the $\delta m_{K(\pi)}$ spectra from simulated decays, 
which include contributions from excited charm baryons and final states with $\taum$ leptons. 
The core of the resolution function has a half-width at half-maximum of about 20\mevcc,
and has a tail toward larger mass (see Appendix). 
The obtained shape parameters are fixed in the fits to data. 

\begin{figure}[tb]
\centering
\includegraphics[width=0.48\textwidth]{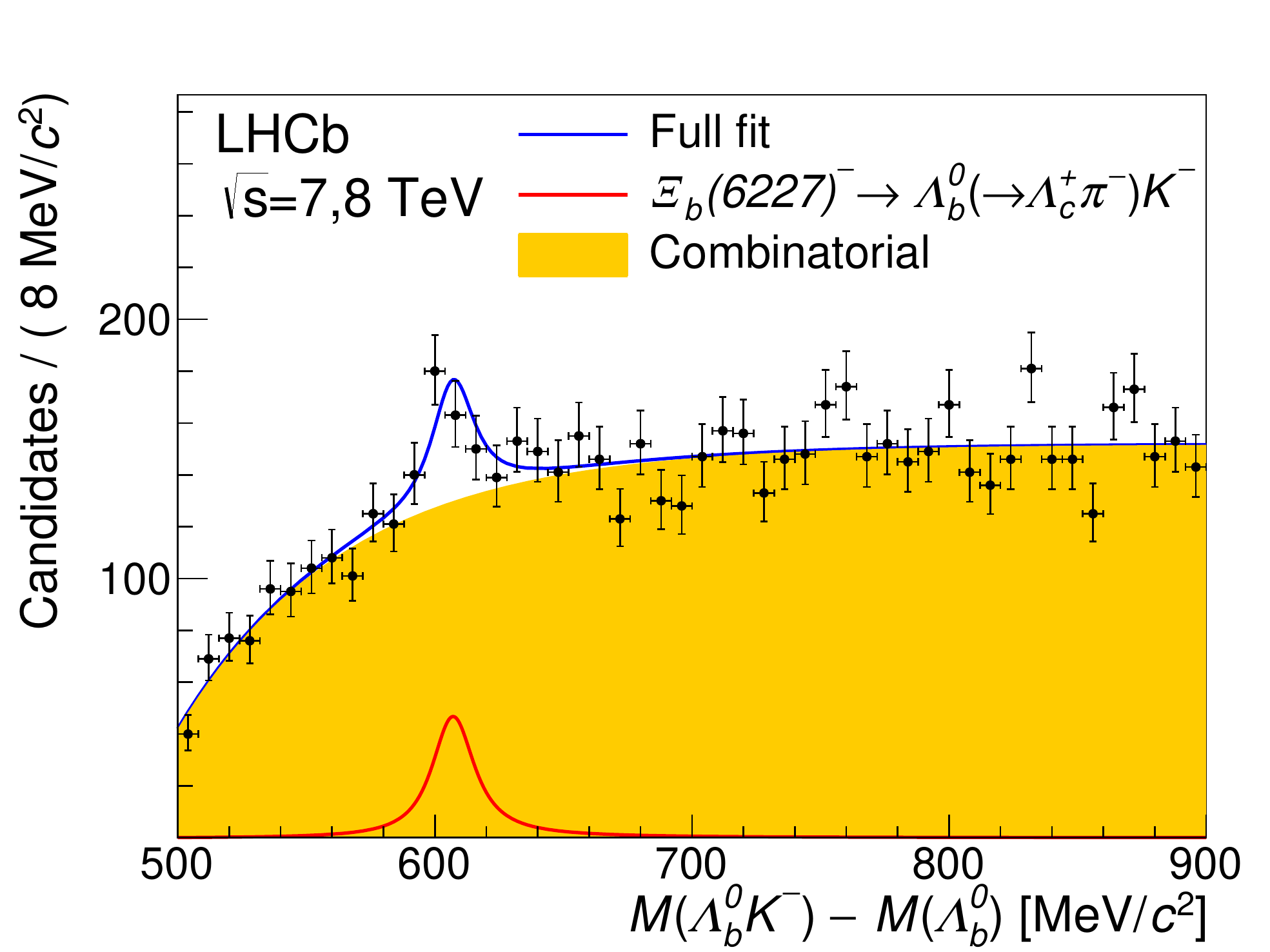}
\includegraphics[width=0.48\textwidth]{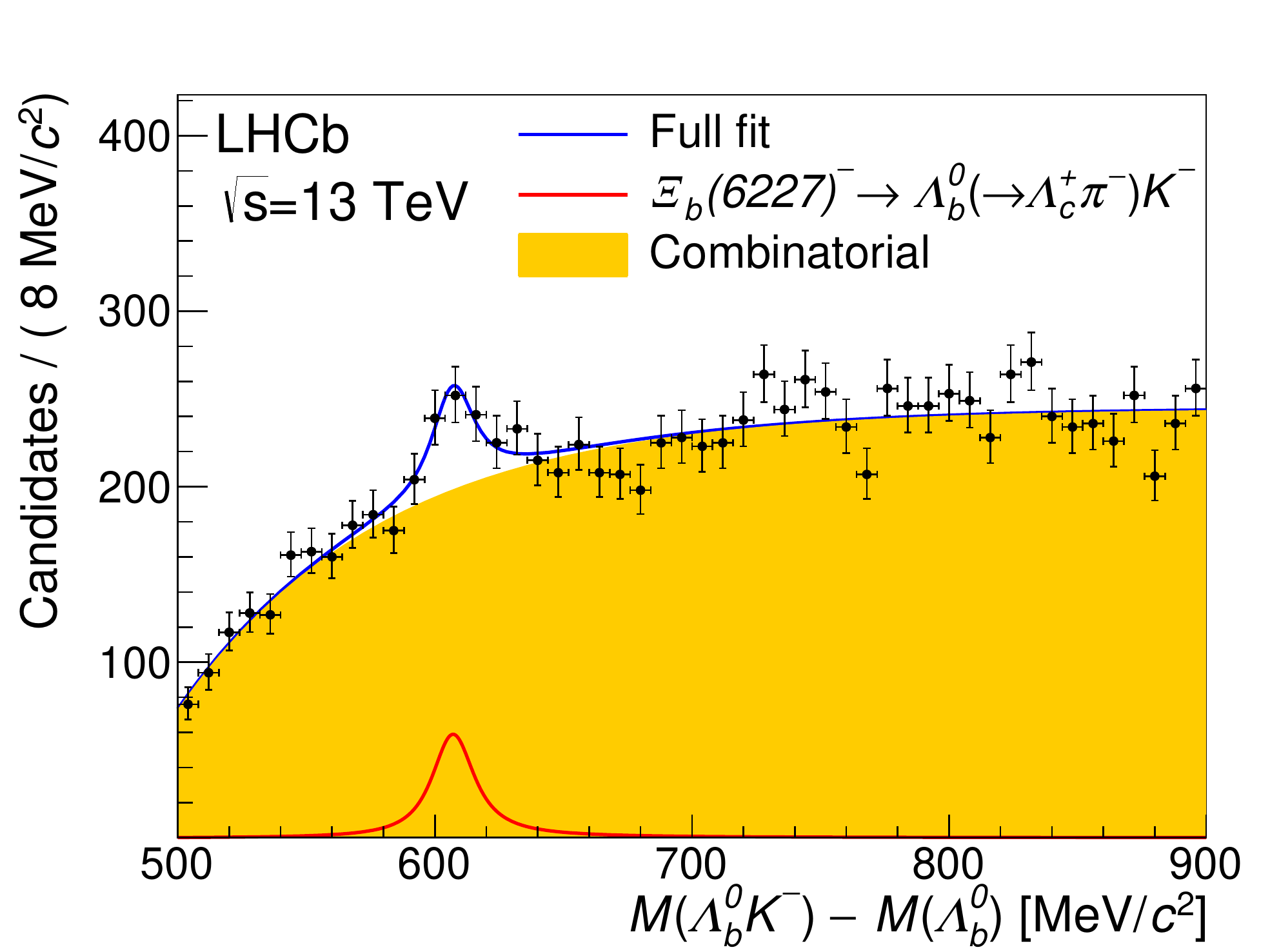}
\includegraphics[width=0.48\textwidth]{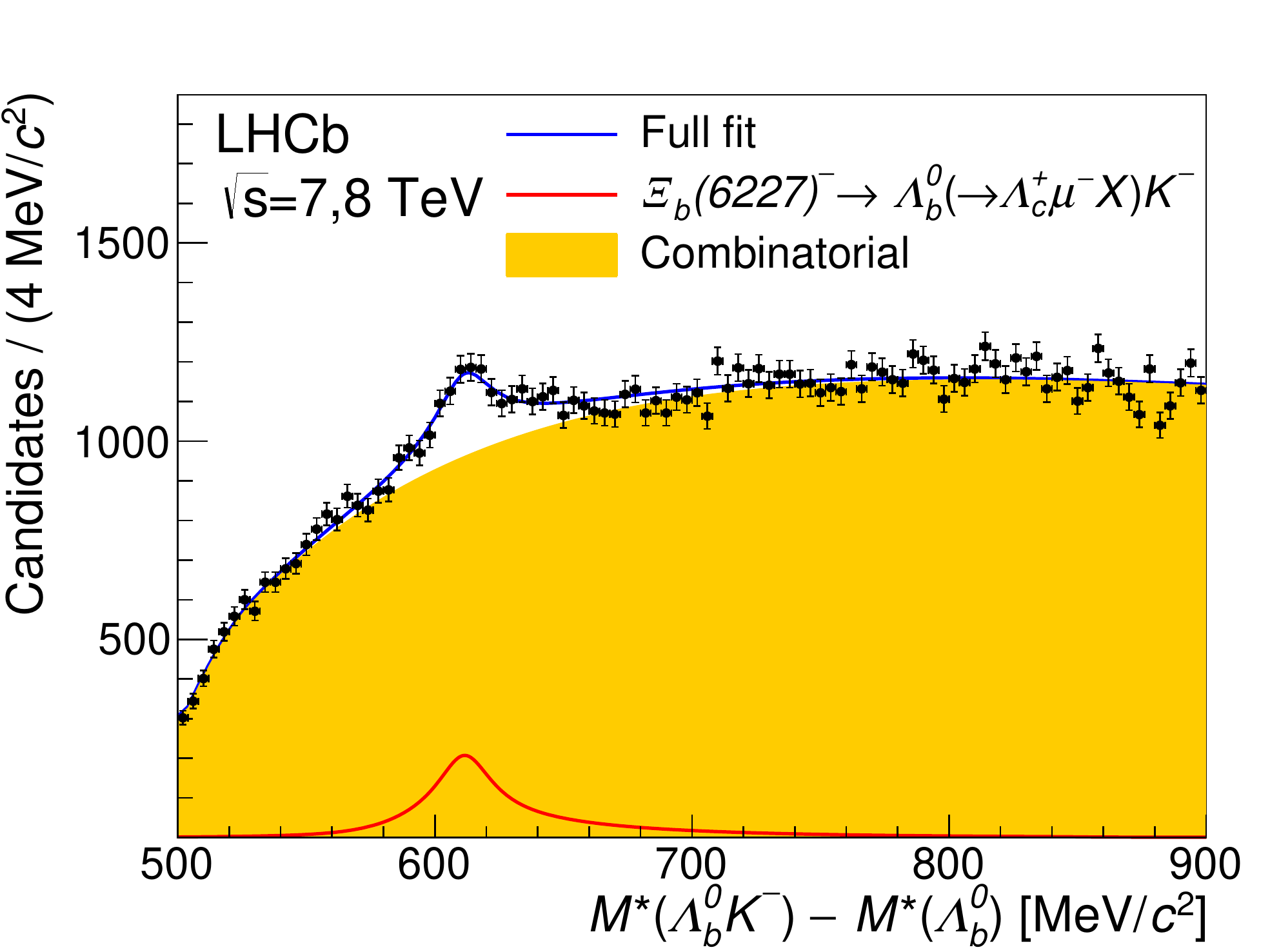}
\includegraphics[width=0.48\textwidth]{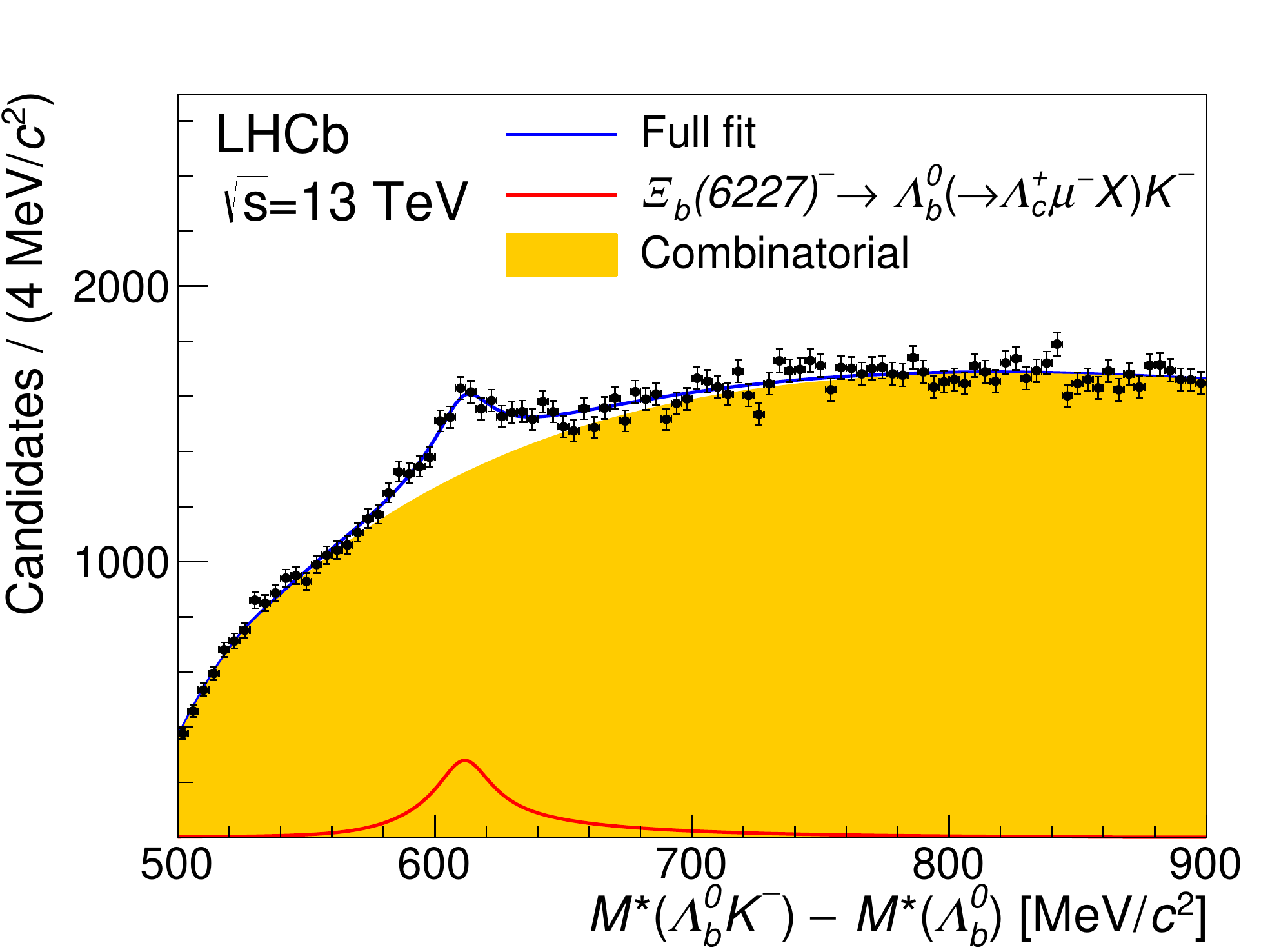}
\includegraphics[width=0.48\textwidth]{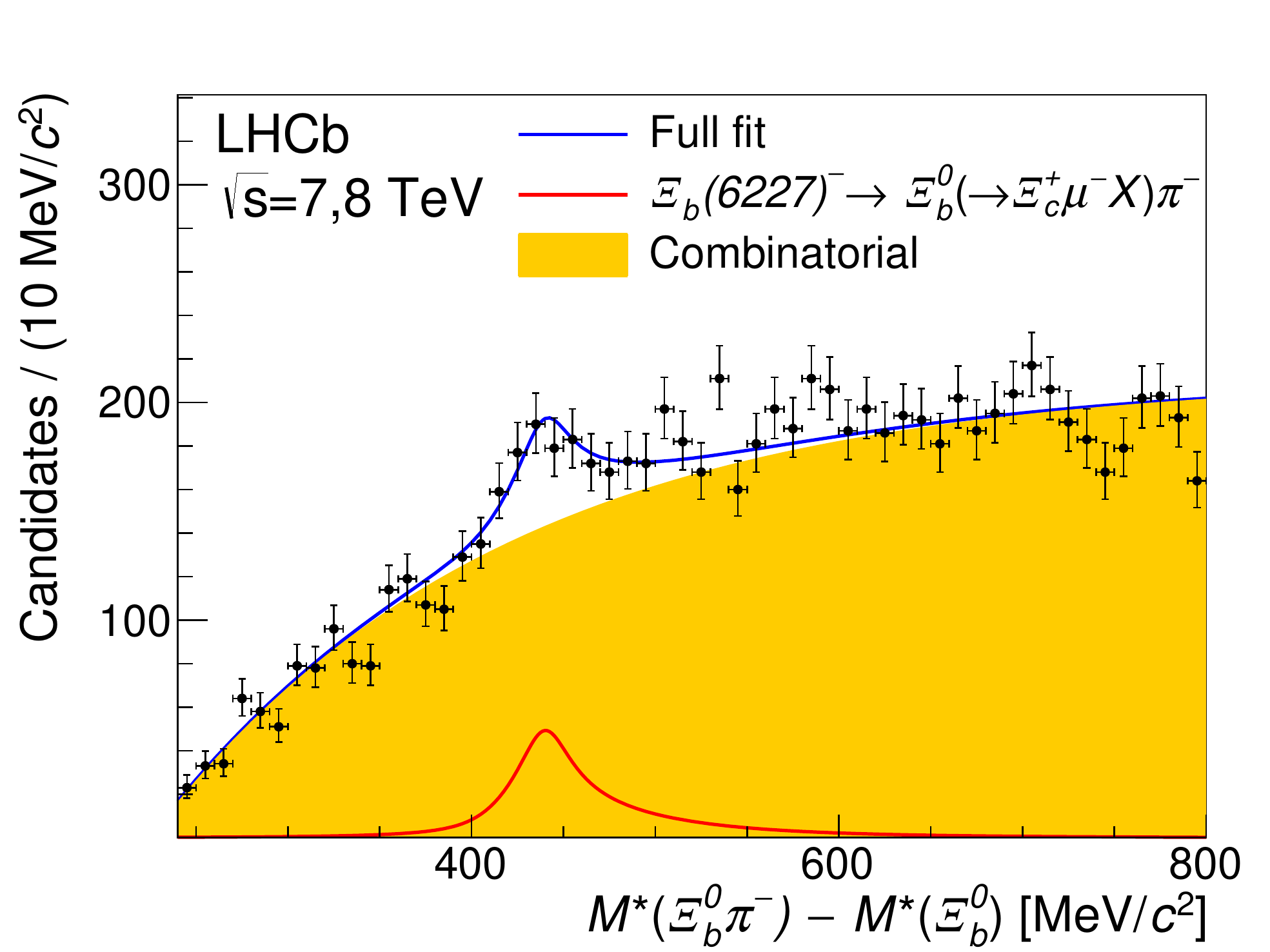}
\includegraphics[width=0.48\textwidth]{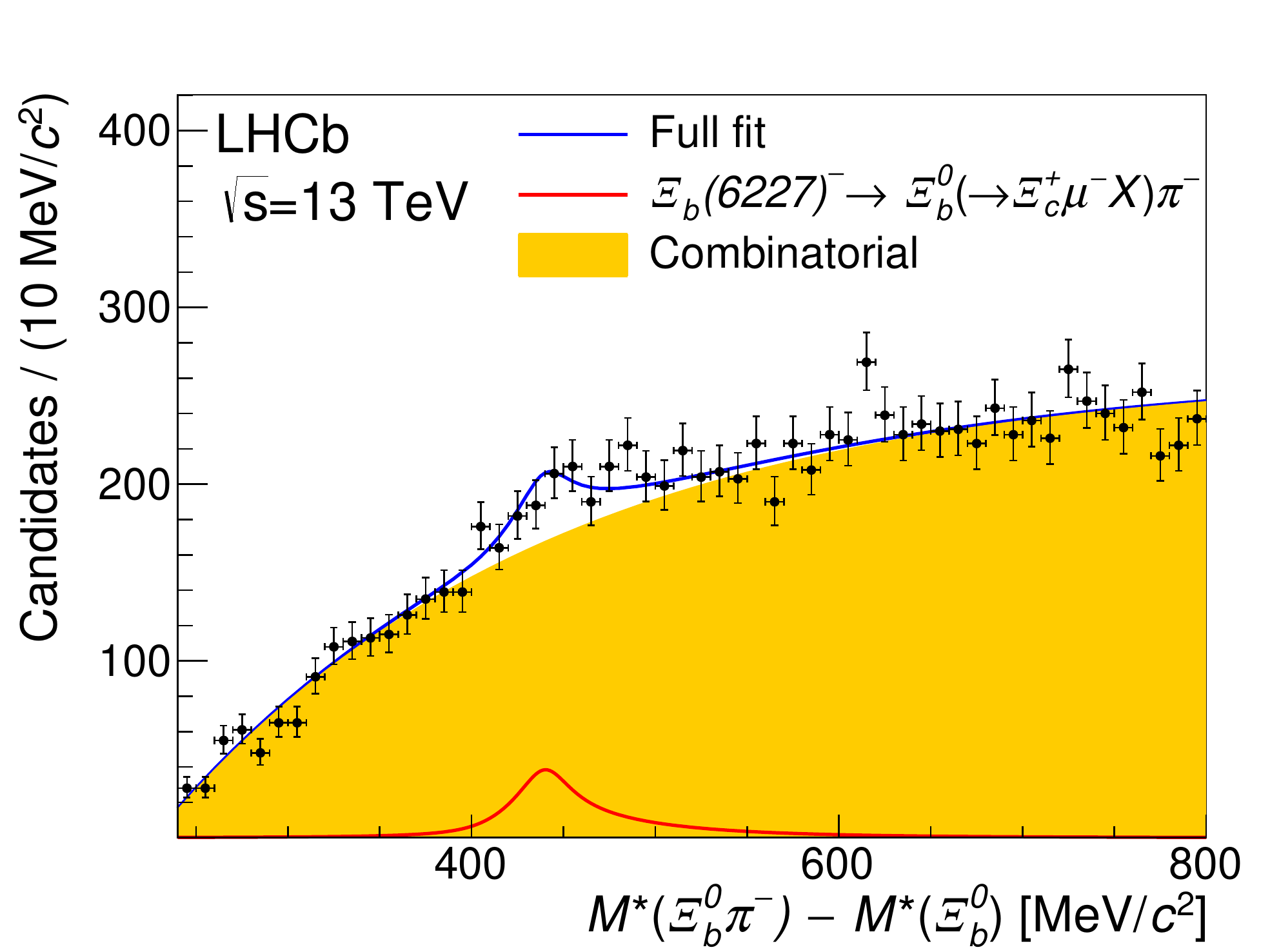}
\caption{\small{Spectra of mass differences for $\XibStar$ candidates, reconstructed in the final states
(top) $\Lb\Km$, with $\Lb\to\Lc\pim$, (middle) $\Lb\Km$, with $\Lb\to\Lc\mun X$,
and (bottom) $\Xibz\pim$, with $\Xibz\to\Xicp\mun X$, along with the results of the fits. 
The left column is for 7,\,8~TeV and the right is for 13~TeV data. The symbol $M^*$ represents the mass after the
constraint $(p_{H_c^+}+p_{\mun}+p_{\rm miss})^2=m_{H_b^0}^2$ is applied, as described in the text.}}
\label{fig:SigModeMassDistributions}
\end{figure}

The $\delta m_K$ and $\delta m_{\pi}$ spectra in data are shown in 
Fig.~\ref{fig:SigModeMassDistributions}.
The $\XibStar$ mass and width are obtained from a simultaneous unbinned maximum-likelihood
fit to the $\delta m_K$ spectra in 7, 8 and 13\tev data, using the $\Lb\to\Lc\pim$ mode.
The signal shape is described by a $P$-wave relativistic Breit-Wigner function~\cite{Jackson:1964zd} with a Blatt-Weisskopf 
barrier factor~\cite{Blatt}, convoluted with a
Gaussian resolution function of width 2.4\mevcc. The mass and width are common parameters in the fit.
The background shape is described by a smooth threshold function~\cite{Brun:1997pa} with shape parameters that are freely and 
independently varied in the fits to the two data sets. A peak is observed in both data sets, with a mean 
$\delta m_K^{\rm peak}=607.3\pm2.0$\mevcc and width $\Gamma_{\XibStar}=18.1\pm5.4$\mevcc.
The peak has a local statistical significance of about 7.9$\sigma$ for the combined fit, based on the difference in 
log-likelihoods between a fit with zero signal and the best fit. The signal yields are given in Table~\ref{tab:sumYield}.

The $\XibStar\to\Lb\Km$ decay with $\Lb\to\Lc\mun X$ is fit in a similar way, except for the different resolution function
(see Appendix). A Gaussian constraint on the width of $\Gamma_{\XibStar}=18.1\pm5.4$\mevcc is applied, 
as obtained from the fit to the hadronic mode, and the mean is freely varied. A peak is observed at a mass difference of
$610.8\pm1.0\stat$\mevcc, which is consistent with that of the hadronic mode, and it contains a yield
about 15 times larger, as expected.
The statistical significance of this peak is about 25$\sigma$, thus clearly establishing this peaking structure.

The $\Xibz\pim$ final state is investigated by examining the $\delta m_{\pi}$ spectra in ${\XibStar\to\Xibz\pim}$ candidate decays,
as shown in the bottom row of Fig.~\ref{fig:SigModeMassDistributions}. The fit is performed in an analogous way to the $\delta m_K$
spectra, except for a different resolution function (see Appendix for $\delta m_{\pi}$ resolution).
The fitted mean of $440\pm5\mevcc$ is consistent with the value expected from the hadronic mode of 
$\delta m_K^{\rm peak}+m_{\Lb}-m_{\Xibz}=435\pm2\mevcc$. The statistical significance of the peak is 9.2$\sigma$.

The production ratios are computed using
\begin{align}
R(\Lb\Km) =\frac{N(\XibStar\to\Lb\Km)}{\epsilon_{\rm rel}N(\Lb)} \kappa\,, \\
R(\Xibz\pim) = \frac{N(\XibStar\to\Xibz\pim)}{\epsilon_{\rm rel}^{\prime}N(\Xibz)}\kappa^{\prime}\,,
\end{align}
\noindent where $N$ represents the yields in Table~\ref{tab:sumYield}, and $\epsilon_{\rm rel}^{(\prime)}$
is the relative efficiency between the $\XibStar$ and $H_b^0$ selections, reported in Table~\ref{tab:seleff}.
\begin{table*}[tb]
\begin{center}
\caption{\small{Relative efficiencies ($\epsilon_{\rm rel}^{(\prime)}$) for the SL modes.
Uncertainties are due only to the finite size of the simulated samples.}}
\begin{tabular}{lcc}
\hline\hline
    Final state          &     $7,\,8$\tev     &       13\tev      \\
\hline
$\Lb\Km$        & $0.295\pm0.006$   &     $0.305\pm0.005$   \\
$\Xibz\pim$     &  $0.236\pm0.007$  &     $0.277\pm0.006$    \\
\hline\hline
\end{tabular}
\label{tab:seleff}
\end{center}
\end{table*}
The quantity $\kappa^{(\prime)}$ represents corrections to the $N(H_b^0)$ SL signal yields 
to account for (i) random $H_c^+\mun$ combinations,
(ii) cross-feed from $\Xibm\to\Xicp\mun X$ decays into the $\Xibz\to\Xicp\mun$ sample, and (iii) slightly different
integrated luminosities used for the $\XibStar$ and $H_b^0$ samples. The contribution from random $H_c^+\mun$ combinations
is estimated from a study of the
wrong-sign ($H_c^+\mup$) and right-sign ($H_c^+\mun$) yields, from which a correction of $1.010\pm0.002$ to both
$R(\Xibz\pim)$ and $R(\Lb\Km)$ is found.
Cross-feeds from SL $\Xibm$ decays, which must be subtracted from $N(\Xibz)$, are inferred by adding a $\pim$ meson to the $\Xicp\mun$ candidate 
and searching for excited $\Xicz$ states. Mass peaks associated with the $\Xic(2645)^0$ and  $\Xic(2790)^0$ resonances are observed, 
although for the former about half is due to $\Xic(2815)^+\to\Xic(2645)^0\pip$ decays, as determined through a study of the $\Xicp\pip$ mass spectrum.
Since the $\Xic(2815)^+\mun$ final state is predominantly from $\Xibz$ decays, this contribution is not subtracted. 
After correcting for the pion detection efficiency, we estimate that $R(\Xibz\pim)$ must be corrected by $1.11\pm0.03$. 
Slightly different-size data samples are used for the $\XibStar$ and inclusive $H_b^0$ yield determinations, which amounts to 
corrections of less than~3\%.

A number of sources of systematic uncertainty have been considered.
For the mass and width, the momentum scale uncertainty of 0.03\%~\cite{LHCb-PAPER-2013-011} leads to a 0.1\mevcc uncertainty 
on $\delta m_{K}$. A fit bias on the mass of 0.1\mevcc is observed in simulation, and is corrected for and
a systematic uncertainty of equal size is assigned. Uncertainty due to
the signal shape model is estimated by using a nonrelativistic Breit-Wigner signal shape and varying the Gaussian resolution by $\pm$10\%
about its nominal value. With these variations, systematic uncertainties of 0.2\mevcc on $\delta m_K$, and 0.9\mevcc on $\Gamma_{\XibStar}$
are obtained. Sensitivity to the background function is assessed by varying the fit range by 100\mevcc on both 
ends, from which maximum
shifts of 0.2\mevcc in the mass and 1.6\mevcc in the width are observed; these values are assigned as systematic uncertainties. 
Adding these systematic uncertainties in quadrature, leads to a total
systematic uncertainty of 0.3\mevcc on the mass and 1.8\mevcc on the width.

The systematic uncertainties affecting the production ratio measurements are listed in Table~\ref{tab:systProd}.
The background shape affects the yield determination, and the associated systematic uncertainty is estimated by varying the 
fit range as described above. (Different background models give smaller deviations.)
For the signal shape, the uncertainty is dominated by the resolution function. In an alternative fit, 
the resolution parameters are allowed to vary within twice the expected uncertainty and we take the difference with respect to the nominal
result as the uncertainty. To assess the dependence on the kinematical properties of the $\XibStar$ resonance, 
the $\pt$ spectrum in simulation is weighted by $1\pm0.01\times p_{\rm T}^{\XibStar}\!/(\gevc)$, based on previous studies 
of the $\Xibz$ and $\Lb$ production spectra~\cite{LHCb-PAPER-2014-021}; the 
relative change in efficiency is assigned as a systematic uncertainty.
The charged-particle tracking efficiency, obtained using large samples of $\jpsi\to\mup\mun$ decays~\cite{LHCb-DP-2013-002}, 
contributes an uncertainty of 1\% to $\epsilon_{\rm rel}^{(\prime)}$. The systematic uncertainty of the PID requirement 
on the $\Km$ or $\pim$ from the $\XibStar$ baryon is determined by comparing the PID response of kaons and pions in 
the $\Lc\to p\Km\pip$ decay between data and simulation, where the latter are
obtained from calibration data, as described previously. 
The uncertainty on $N(H_b^0)$ is taken as the quadratic sum of the uncertainties on the fitted yields and the uncertainties on the 
$\kappa^{(\prime)}$ corrections. Lastly, the finite size of the simulated samples is taken into account.
\begin{table*}[tb]
\begin{center}
\caption{\small{Summary of systematic uncertainties on $R(\Lb\Km)$ and $R(\Xibz\pim)$, in units of $10^{-3}$.}}
\begin{tabular}{lcccc}
\hline\hline
                  & \multicolumn{2}{c}{$R(\Lb\Km)~[10^{-3}]$\rule{0pt}{2.4ex}} & \multicolumn{2}{c}{$R(\Xibz\pim)~[10^{-3}]$} \\
Source            &      $7,\,8$\tev    &     13\tev                &      $7\,,8$\tev    &     13\tev    \\
\hline         
Background shape  &        0.3     &     0.3                &        6.0     &     3.0 \\
Signal shape      &        0.1     &     0.1                &        1.0     &     0.2 \\
$\XibStar$ $\pt$  &        $^{+0.16}_{-0.27}$ & $^{+0.14}_{-0.33}$ &       $^{+2.5}_{-3.2}$ & $^{+0.9}_{-1.5}$ \\ 
Tracking efficiency    &    0.03    &     0.03                &       0.5     &     0.2 \\
PID requirement    &        0.05    &     0.06                &       0.5     &     0.2 \\
$N(H_b^0)$            &      0.01      &   0.01                &        1.4     &     0.7  \\
Simulated sample size &        0.07    &     0.05                &        1.4     &     0.6 \\
\hline
Total        &     0.4 &      0.4                  &        7.0     &     3.3 \\
\hline\hline
\end{tabular}
\label{tab:systProd}
\end{center}
\end{table*}

In summary, we report the first observation of a new state, assumed to be an excited $\Xibm$ state, using 
$pp$ collision data samples collected by LHCb at $\sqrt{s}=7\,,8$ and 13\tev. The mass and width are
measured to be
\begin{align*}
m_{\XibStar}-m_{\Lb} &= 607.3\pm2.0\stat\pm0.3\syst \mevcc, \\
\Gamma_{\XibStar} &= 18.1\pm5.4\stat\pm1.8\syst \mevcc, \\
m_{\XibStar} &= 6226.9\pm2.0\stat\pm0.3\syst\pm0.2(\Lb) \mevcc,
\end{align*}
\noindent where for the last result we have used $m_{\Lb}=5619.58\pm0.17\mevcc$~\cite{PDG2017}.

We have also measured the relative production rates to two final states, $\Lb\Km$ and $\Xibz\pim$, as summarized in
Table~\ref{tab:RvaluesSum}. The $R(\Lb\Km)$ values from the hadronic mode are consistent with those obtained in the SL mode,
and are about an order of magnitude smaller than $R(\Xibz\pim)$. Assuming 
$f_{\Xibz}\simeq0.1 f_{\Lb}$~\cite{voloshin,Hsiao:2015txa,Jiang:2018iqa}, we find that the ratio of branching fractions
$\BR(\XibStar\to\Lb\Km)/\BR(\XibStar\to\Xibz\pim)\simeq 1$, albeit with sizable uncertainty ($\approx\pm0.5$) due to 
theoretical assumptions and the values of experimental inputs.

\begin{table*}[tb]
\begin{center}
\caption{\small{Measured ratios $R(\Lb\Km)$ and $R(\Xibz\pim)$ for $7\,,8$ and 13~TeV data, in units of $10^{-3}$. 
The uncertainties are statistical (first) and systematic (second).}}
\begin{tabular}{lcc}
\hline\hline
Quantity~$[10^{-3}]$\rule{0pt}{2.4ex}       &    $7\,,8$\tev       &   13\tev \\
\hline
$R(\Lb\Km)$\rule{0pt}{2.4ex}    & $3.0\pm0.3\pm0.4$  &  $3.4\pm0.3\pm0.4$   \\
$R(\Xib\pim)$                   & $47\pm10\pm7$      &  $22\pm6\pm3$      \\
\hline\hline
\end{tabular}
\label{tab:RvaluesSum}
\end{center}
\end{table*}
The mass of this structure and the observed decay modes are consistent with expectations of either a
$\Xib(1P)^-$ or $\Xib(2S)^-$ state~\cite{Ebert:2011kk,Ebert:2007nw,Roberts:2007ni,Garcilazo:2007eh,Chen:2014nyo,Mao:2015gya,Grach:2008ij,PhysRevD.87.034032,Karliner:2008sv,Wang:2010it,Valcarce:2008dr,Vijande:2012mk,Wang:2017kfr,Wang:2017goq,Chen:2016phw,Thakkar:2016dna}. 
As there are several excited $\Xibm$ states expected in this mass region, the presence of more than one of these states 
contributing to this peak cannot be excluded.
More precise measurements of the width and the relative branching 
fractions to $\Lb\Km$ and $\Xibz\pim$, as well as $\Xib^{\prime}\pim$ and $\Xib^*\pim$, could help to determine the 
$J^P$ quantum numbers of this state~\cite{Wang:2017kfr}.


\section*{Acknowledgements}
%
%
\noindent We express our gratitude to our colleagues in the CERN
accelerator departments for the excellent performance of the LHC. We
thank the technical and administrative staff at the LHCb
institutes. We acknowledge support from CERN and from the national
agencies: CAPES, CNPq, FAPERJ and FINEP (Brazil); MOST and NSFC
(China); CNRS/IN2P3 (France); BMBF, DFG and MPG (Germany); INFN
(Italy); NWO (The Netherlands); MNiSW and NCN (Poland); MEN/IFA
(Romania); MinES and FASO (Russia); MinECo (Spain); SNSF and SER
(Switzerland); NASU (Ukraine); STFC (United Kingdom); NSF (USA).  We
acknowledge the computing resources that are provided by CERN, IN2P3
(France), KIT and DESY (Germany), INFN (Italy), SURF (The
Netherlands), PIC (Spain), GridPP (United Kingdom), RRCKI and Yandex
LLC (Russia), CSCS (Switzerland), IFIN-HH (Romania), CBPF (Brazil),
PL-GRID (Poland) and OSC (USA). We are indebted to the communities
behind the multiple open-source software packages on which we depend.
Individual groups or members have received support from AvH Foundation
(Germany), EPLANET, Marie Sk\l{}odowska-Curie Actions and ERC
(European Union), ANR, Labex P2IO and OCEVU, and R\'{e}gion
Auvergne-Rh\^{o}ne-Alpes (France), Key Research Program of Frontier
Sciences of CAS, CAS PIFI, and the Thousand Talents Program (China),
RFBR, RSF and Yandex LLC (Russia), GVA, XuntaGal and GENCAT (Spain),
Herchel Smith Fund, the Royal Society, the English-Speaking Union and
the Leverhulme Trust (United Kingdom).


\clearpage

{\noindent\normalfont\bfseries\Large Appendix}
\appendix
\vspace{0.5in}

The mass resolution functions for the ${\XibStar\to\Lb\Km}$ and
${\XibStar\to\Xibz\pim}$ semileptonic decays are provided below.

\begin{figure}[thb]
\centering
\includegraphics[width=0.49\textwidth]{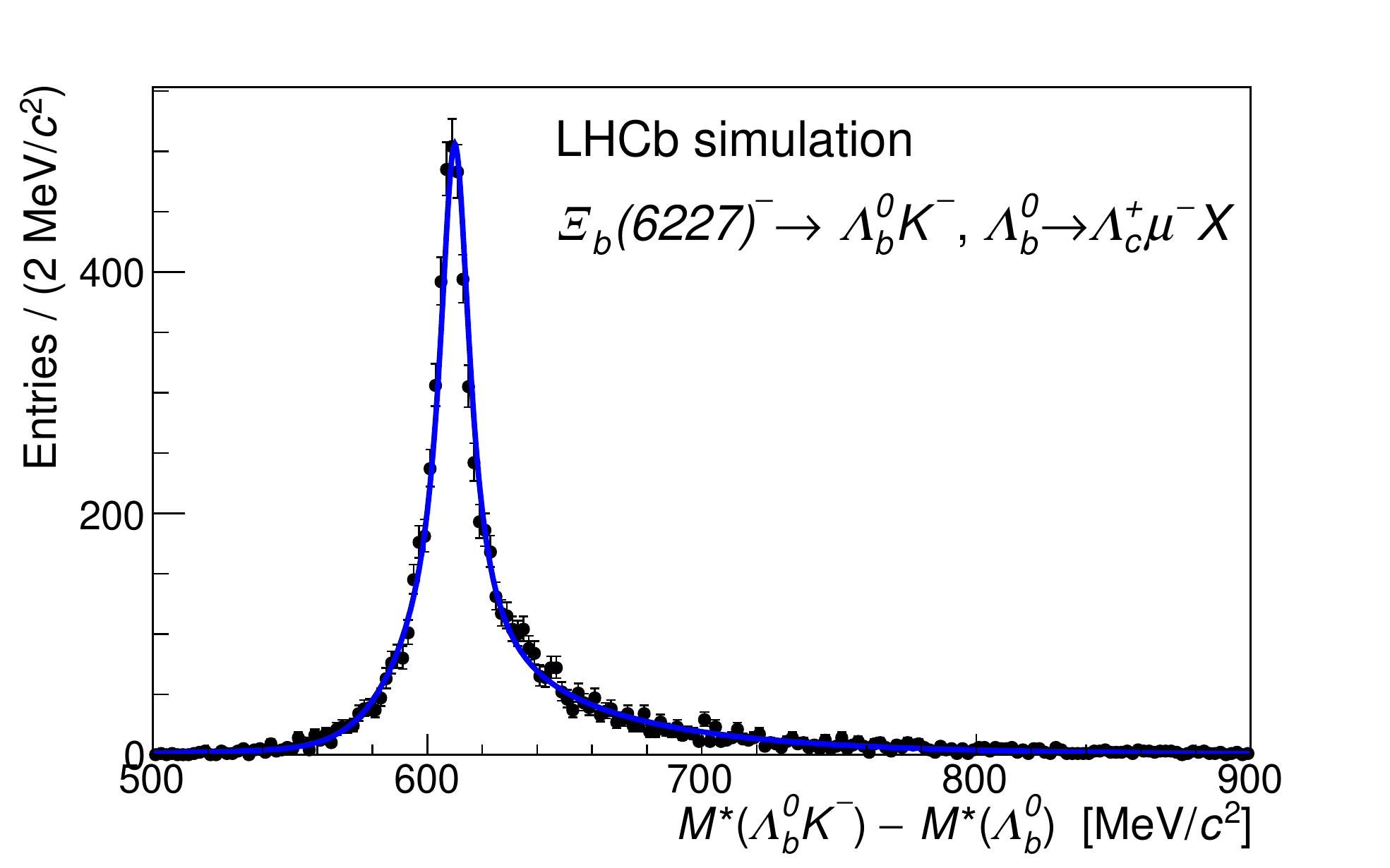}
\includegraphics[width=0.49\textwidth]{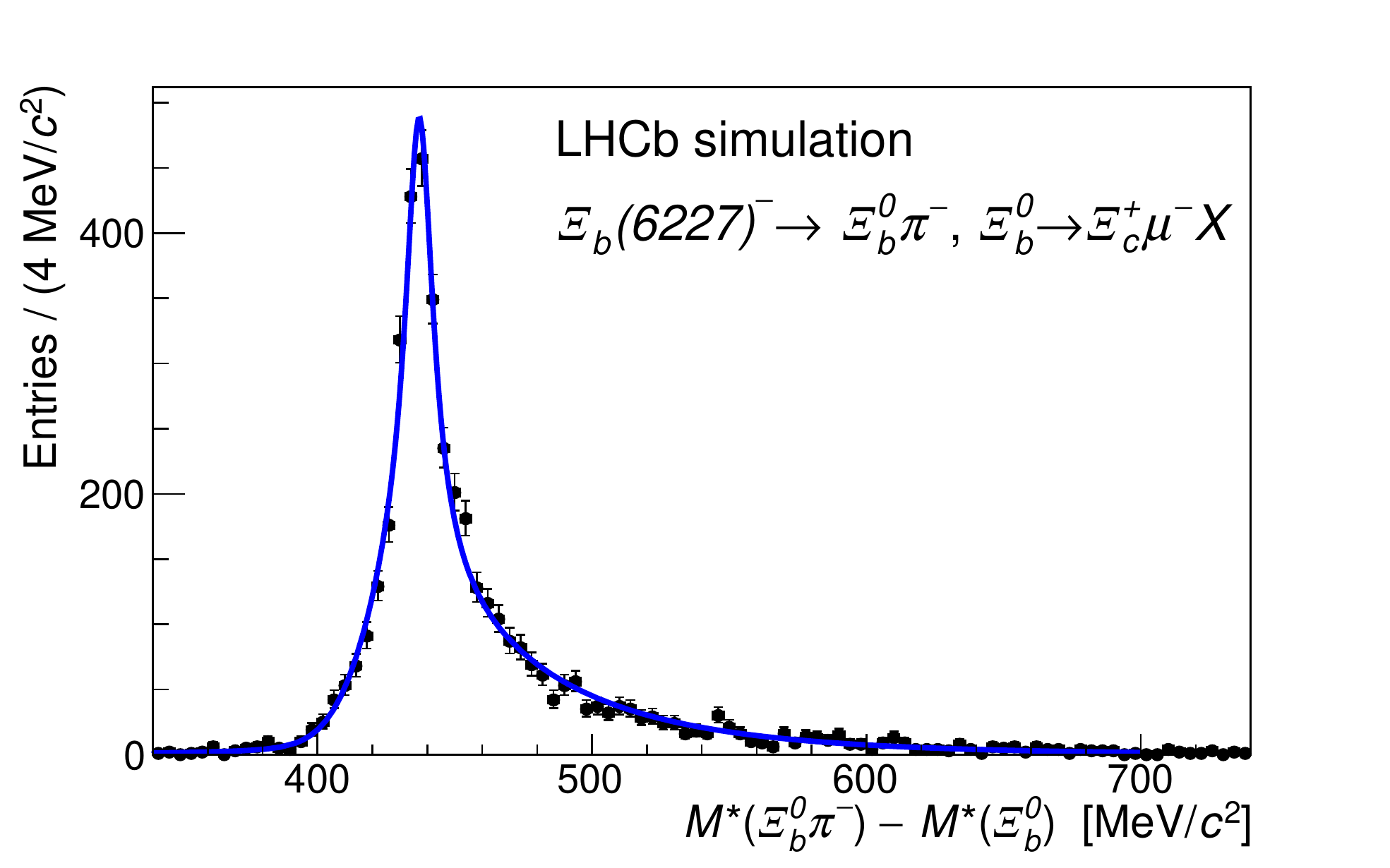}
\caption{\small{Distribution of (left) $M^*(\Lb\Km)-M^*(\Lb)$ 
for simulated  $\XibStar\to\Lb\Km$ decays, where $\Lb\to\Lc\mun X$, and 
(right) $M^*(\Xibz\pim)-M^*(\Xibz)$ 
for simulated $\XibStar\to\Xibz\pim$ decays, where $\Xibz\to\Xicp\mun X$.
The symbol $M^*$ represents the mass after the
constraint $(p_{H_c^+}+p_{\mun}+p_{\rm miss})^2=m_{H_b^0}^2$ is applied, as described in the text.
The natural width used in the simulation is set to a negligible value, so that
these spectra are due entirely to the mass resolution.
Fits to the sum of a nonrelativistic Breit-Wigner function and a Crystal Ball 
function~\cite{Skwarnicki:1986xj} with a common mean value are overlaid.}}
\label{fig:ResolutionFunction}
\end{figure}


\clearpage
\addcontentsline{toc}{section}{References}
\setboolean{inbibliography}{true}
\bibliographystyle{LHCb}
\bibliography{main,LHCb-PAPER,LHCb-CONF,LHCb-DP,LHCb-TDR,stevesRefs}



 
\newpage
\centerline{\large\bf LHCb collaboration}
\begin{flushleft}
\small
R.~Aaij$^{27}$,
B.~Adeva$^{41}$,
M.~Adinolfi$^{48}$,
C.A.~Aidala$^{73}$,
Z.~Ajaltouni$^{5}$,
S.~Akar$^{59}$,
P.~Albicocco$^{18}$,
J.~Albrecht$^{10}$,
F.~Alessio$^{42}$,
M.~Alexander$^{53}$
% ...

{\footnotesize \it
% ...
$ ^{5}$Clermont Universit{\'e}, Universit{\'e} Blaise Pascal, CNRS/IN2P3, LPC, Clermont-Ferrand, France\\
% ...
$ ^{10}$Fakult{\"a}t Physik, Technische Universit{\"a}t Dortmund, Dortmund, Germany\\
% ...
$ ^{18}$INFN Laboratori Nazionali di Frascati, Frascati, Italy\\
% ...
$ ^{27}$Nikhef National Institute for Subatomic Physics, Amsterdam, Netherlands\\
% ...
$ ^{41}$Instituto Galego de F{\'\i}sica de Altas Enerx{\'\i}as (IGFAE), Universidade de Santiago de Compostela, Santiago de Compostela, Spain\\
$ ^{42}$European Organization for Nuclear Research (CERN), Geneva, Switzerland\\
% ...
$ ^{48}$H.H. Wills Physics Laboratory, University of Bristol, Bristol, United Kingdom\\
% ...
$ ^{53}$School of Physics and Astronomy, University of Glasgow, Glasgow, United Kingdom\\
% ...
$ ^{59}$University of Cincinnati, Cincinnati, OH, United States\\
% ...
$ ^{61}$Syracuse University, Syracuse, NY, United States\\
% ...
$ ^{73}$University of Michigan, Ann Arbor, United States, associated to $^{61}$\\
% ...
}
\end{flushleft}

\end{document}